%% file: gatepmb.tex
\documentclass[12pt|a4paper]{iopart}
\include{gate_newcommand}

\usepackage{html}
\usepackage{graphicx}
\usepackage{ulem}

\begin{document}

\title[\gate: a simulation toolkit for PET and SPECT]{\gate: a simulation toolkit for PET and SPECT}

\author{S.~Jan$^{1}$,
G.~Santin$^{2}$\footnote[3]{Present address: ESA - European Space and Technology Centre, 2200 AG Noordwijk, The Netherlands},
D.~Strul$^{2}$\footnote[5]{Present address: CNRS-LESIA, Observatoire de Meudon, F-92190 Meudon, France},
S.~Staelens$^{3}$,
K.~Assi\'e$^{4}$,
D.~Autret$^{5}$,
S.~Avner$^{6}$,
R.~Barbier$^{7}$,
M.~Bardi\`es$^{5}$,
P.~M.~Bloomfield$^{8}$,
D.~Brasse$^{6}$,
V.~Breton$^{9}$,
P.~Bruyndonckx$^{10}$,
I.~Buvat$^{4}$,
A.~F.~Chatziioannou$^{11}$,
Y.~Choi$^{12}$,
Y.~H.~Chung$^{12}$,
C.~Comtat$^{1}$,
D.~Donnarieix$^{9,21}$,
L.~Ferrer$^{5}$,
S.~J.~Glick$^{13}$,
C.~J.~Groiselle$^{13}$,
D.~Guez$^{14}$,
P.-F.~Honore$^{14}$,
S.~Kerhoas-Cavata$^{14}$,
A.~S.~Kirov$^{15}$,
V.~Kohli$^{11}$,
M.~Koole$^{3}$,
M.~Krieguer$^{10}$,
D.~J.~van der Laan$^{16}$,
F.~Lamare$^{17}$,
G.~Largeron$^{7}$,
C.~Lartizien$^{18}$,
D.~Lazaro$^{9}$,
M.~C.~Maas$^{16}$,
L.~Maigne$^{9}$,
F.~Mayet$^{19}$,
F.~Melot$^{19}$,
C.~Merheb$^{14}$,
E.~Pennacchio$^{7}$,
J.~Perez$^{20}$,
U.~Pietrzyk$^{20}$,
F.~R.~Rannou$^{11,22}$,
M.~Rey$^{2}$,
D.~R.~Schaart$^{16}$,
C.~R.~Schmidtlein$^{15}$,
L.~Simon$^{2}$\footnote[7]{Present address: Institut Curie, Service de Physique M\'edicale, F-75005 Paris, France},
T.~Y.~Song$^{12}$,
J.-M.~Vieira$^{2}$,
D.~Visvikis$^{17}$,
R.~Van de Walle$^{3}$,
E.~Wie\"ers$^{10,23}$,
and C.~Morel$^{2}$\footnote[8]{Corresponding author : christian.morel@epfl.ch}}
\address{$^1$ Service Hospitalier Fr\'ed\'eric Joliot (SHFJ), CEA, F-91401 Orsay, France}
\address{$^2$ LPHE, Swiss Federal Institute of Technology (EPFL), CH-1015 Lausanne, Switzerland}
\address{$^{3}$ ELIS, Ghent University, B-9000 Ghent, Belgium}
\address{$^4$ INSERM U494, CHU Piti\'e-Salp\^etri\`ere, F-75634 Paris, France }
\address{$^5$ INSERM U601, CHU Nantes, F-44093 Nantes, France}
\address{$^{6}$ Institut de Recherches Subatomiques, CNRS/IN2P3 et Universit\'e Louis Pasteur, F-67037 Strasbourg, France}
\address{$^{7}$ Institut de Physique Nucl\'eaire de Lyon, CNRS/IN2P3 et Universit\'e Claude Bernard, F-69622 Villeurbanne, France}
\address{$^{8}$ PET Group, Centre for Addiction and Mental Health, Toronto, Ontario M5T 1R8, Canada}
\address{$^9$ Laboratoire de Physique Corpusculaire, CNRS/IN2P3, Universit\'e Blaise Pascal, Campus des C\'ezeaux, F-63177 Aubi\`ere, France}
\address{$^{10}$ Inter-University Institute for High Energies, Vrije Universiteit Brussel, B-1050 Brussel, Belgium}
\address{$^{11}$ Crump Institute for Molecular Imaging, University of California, Los Angeles, Califronia 90095-1770, USA}
\address{$^{12}$ Department of Nuclear Medicine, Samsung Medical Center, Sungkyunkwan University School of Medicine, Seoul 135-710,  Korea}
\address{$^{13}$ University of Massachusetts Medical School, Division of Nuclear Medicine, Worcester, MA 01655, USA}
\address{$^{14}$ DAPNIA, CEA Saclay, F-91191 Gif-Sur-Yvette, France}
\address{$^{15}$ Department of Medical Physics, Memorial Sloan-Kettering Cancer Center, New York, NY 10021, USA}
\address{$^{16}$ Delft University of Technology, IRI, Radiation Technology, 2629 JB Delft, The Netherlands}
\address{$^{17}$ INSERM U650, Laboratoire de Traitement de l'Information M\'edicale (LATIM), CHU Morvan, F-29609 Brest, France}
\address{$^{18}$ ANIMAGE-CERMEP, Universit\'e Claude Bernard Lyon 1, F-69003 Lyon, France}
\address{$^{19}$ Laboratoire de Physique Subatomique et de Cosmologie, CNRS/IN2P3 et Universit\'e Joseph Fourier, F-38026 Grenoble, France}
\address{$^{20}$ Institute of Medicine, Forschungszemtrum Juelich, D-52425 Juelich, Germany}
\address{$^{21}$ D\'epartement de Curieth\'erapie-Radioth\'erapie, Centre Jean Perrin, F-63000 Clermont-Ferrand, France}
\address{$^{22}$ Departamento de Ingenieria Informatica, Universidad de Santiago de Chile, Santiago, Chile}
\address{$^{23}$ Nucleair Technologisch Centrum, Dept. Industri\"ele Wetenschappen en Technologie, Hogeschool Limburg, B-3590 Diepenbeek, Belgium}

\begin{abstract}
Monte Carlo simulation is an essential tool in emission tomography that can assist in the design of new medical imaging devices, the optimization of acquisition protocols, and the development or assessment of image reconstruction algorithms and correction techniques. \gate, the~{\it \geant~Application for Tomographic Emission}, encapsulates the \geant~libraries to achieve a modular, versatile, scripted simulation toolkit adapted to the field of nuclear medicine. In particular, \gate~allows the description of time-dependent phenomena such as source or detector movement, and source decay kinetics. This feature makes it possible to simulate time curves under realistic acquisition conditions and to test dynamic reconstruction algorithms. This paper gives a detailed description of the design and development of \gate~by the \OG~collaboration, whose continuing objective is to improve, document, and validate \gate~by simulating commercially available imaging systems for PET and SPECT. Large effort is also invested in the ability and the flexibility to model novel detection systems or systems still under design. A public release of \gate~licensed under the GNU Lesser General Public License can be downloaded at the address http://www-lphe.epfl.ch/GATE/. Two benchmarks developed for PET and SPECT to test the installation of \gate~and to serve as a tutorial for the users are presented. Extensive validation of the \gate~simulation platform has been started, comparing simulations and measurements on commercially available acquisition systems. References to those results are listed. The future prospects toward the {\it gridification} of \gate~and its extension to other domains such as dosimetry are also discussed.
\end{abstract}


\submitto{\PMB}

\maketitle

\input{pmb.sec.1}
\input{pmb.sec.3}

\input{pmb.sec.6}
\input{pmb.sec.7}
\input{pmb.sec.9}

\input{pmb.sec.10}

\input{pmb.sec.11}
\input{pmb.sec.12}

\section*{Acknowledgments}

This work was supported by the Swiss National Science Foundation under Grant Nos. 2153-063870 and 205320-100472, by the Korea Health 21 R\&D Project under Grant No. 02-PJ3-PG6-EV06-0002, by the US National Cancer Institute under Grant Nos. CA-78573 and CA059017-12, by the Agassiz Foundation, by the Swiss-French Integrated Action Programme Germaine de Sta\"el, by the Institute for the Promotion of Innovation by Science and Technology in Flanders (IWT, Belgium), and by the Fund for Scientific Research in Flanders (FWO, Belgium).

The \OG~collaboration is grateful to the \geant~collaboration, and particularly to the Low Energy Working Group of the \geant~collaboration for their help and support during the starting phase of the development of \gate. In addition, U494 is grateful to the Ligue Nationale Contre la Cancer for supporting Karine Assi\'e's research. The authors also appreciate the support provided by Ghent University and by Philips Medical Systems (INSERM U650). 

\input{biblio}

\end{document}

%% file: gate_newcommand.tex
\newcommand{\gate}      {{GATE}}

\newcommand{\OG}    {{OpenGATE}}
\newcommand{\geant}     {{Geant4}}
\newcommand{\eg}        {{\it e.g.}}
\newcommand{\ie}        {{\it i.e.}}
\newcommand{\dig}       {{DigiGATE}}
\newcommand{\rt}       {{\it ROOT}}

%% file: pmb.sec.1.tex
\section{Introduction}
\label{sec:intro}

%
%

Emission tomography is becoming increasingly important in modern medicine for both diagnostic and treatment monitoring, with a demand for higher imaging quality, accuracy, and speed. Recently enhanced by the wider availability of powerful computer clusters, Monte Carlo simulations have become an essential tool for current and future emission tomography development. Examples of research areas benefiting from these developments are the design of new medical imaging devices, the optimization of acquisition protocols, and the development and assessment of image reconstruction algorithms and correction techniques.

Currently there are numerous Monte Carlo simulation packages for either PET (Positron Emission Tomography) or SPECT (Single Photon Emission Computer Tomography), each with different advantages, disadvantages, and levels of reliability (Buvat and Castiglioni 2002). Accurate and versatile {\it general-purpose} simulation packages such as Geant3 (Brun \etal 1987), EGS4 (Bielajew \etal 1994), MCNP (Briesmeister 1993), and most recently \geant~(Agnostelli \etal 2003) are available. These packages include well-validated physics models, geometry modeling tools, and efficient visualization utilities. However, it is quite difficult to tailor these packages to PET and SPECT. On the other hand, the dedicated Monte Carlo codes developed for PET and SPECT suffer from a variety of drawbacks and limitations in terms of validation, accuracy and support (Buvat and Castiglioni 2002). While an adaptation of EGS4 to radiation therapy applications exists (Rogers \etal 1995, Kawrakov and Rogers 2003), there are no dedicated PET or SPECT Monte Carlo programs that are detailed and flexible enough for realistic simulations of emission tomography detector geometries. SimSET (Harrison \etal 1993), one of the most powerful dedicated codes enabling PET and SPECT simulations, enables a precise and efficient modelling of physics phenomena and basic detector designs (\eg~ring detectors and planar detectors), but it has limitations with respect to the range of detector geometries that can be modeled. For example, a detector ring can not be subdivided into individual crystals and the gaps between the crystals  and the grouping of crystals into blocks cannot be modeled. In addition, neither SimSET, nor any other publicly available codes account for time explicitly, which limits their use for modeling time dependent processes such as tracer kinetics or bed motion. 

Clearly, a Monte Carlo code capable of accomodating complex scanner geometry and imaging configurations in a user-friendly way, while retaining the comprehensive physics modeling abilities of the general purpose codes is needed. Furthermore, the need is to have a platform that can model decay kinetics, deadtime, and movement, while benefiting from the same versatility and support as that of the general-purpose simulation codes. Object-oriented technology appeared to be the best choice to ensure high modularity and re-usability for a PET and SPECT simulation tool. Therefore, we selected the simulation toolkit developed in C++ by the \geant~collaboration (Agnostelli \etal 2003), and decided to foster long-term support and maintenance by sharing code development among many research groups forming the OpenGATE collaboration.

This paper presents a detailed description of the design and development of a Monte Carlo tool by the \OG~collaboration which satisfies the requirements mentioned above. It was launched at first (Strul 2001, Strul 2001b) as an aid in the design of the ClearPET prototype scanners being developed by the Crystal Clear collaboration (Ziemons \etal 2003). \gate, the {\it \geant~Application for Tomographic Emission} (Santin \etal 2003, Strul \etal 2003, Assi\'e \etal 2004), incorporates the \geant~libraries in a modular, versatile, and scripted simulation toolkit that is specifically adapted to the field of nuclear medicine. A public release of \gate~licensed under the GNU Lesser General Public License ({\it LGPL} n.d.) can be downloaded at the address http://www-lphe.epfl.ch/GATE/.

%% file: pmb.sec.3.tex
\section{\gate~basics}
\label{sec:design}
%
%

\gate~was designed with several objectives in mind. First, the use of the \gate~software should not require any knowledge of C++. End-users from the nuclear medicine community should be able to use \gate~without worrying about the programming details. Second, as many nuclear medicine diagnostic techniques share similar concepts, \gate~software components should be general enough to be reused from one context to another. Last, \gate~should be modular, and thus be able to evolve as new applications are envisioned.

\subsection{Architecture}

The requirements discussed above are met using a layered architecture. The core of \gate, developed in C++, defines the main tools and features of \gate. The application layer is an extensible set of C++ classes based on the \gate~core. On top of the application layer is the user layer, where end-users can simulate experiments using an extended version of the \geant~scripting language.

\begin{figure}[h]
\begin{center}
\includegraphics[scale=0.4,angle=0]{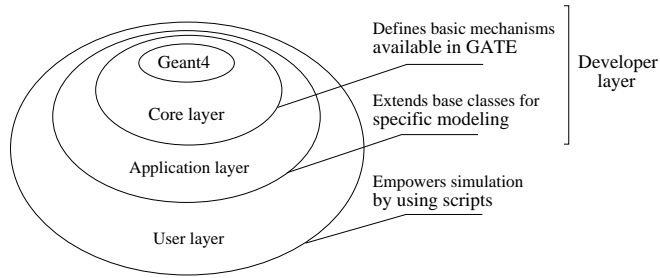}
\caption{Sketch of the layered architecture of GATE.}
\label{fig:GATEarchi}
\end{center}
\end{figure}

The \gate~developer layer consists of the {\it core layer} and the {\it application layer}. It is built from the various classes that provide the most general features of \gate. These classes define which tools are available, what developers can do, and how they can do it. The core layer includes some base classes that are common or even mandatory in all \geant-based simulations, such as those involved in the construction of the geometry, the interaction physics, the event generation, and the visualization management. In addition, the core layer includes classes that are specific to \gate~simulations, such as the \gate~virtual clock for time management. Thus, the core layer defines the basic mechanisms available in \gate~for geometry definition, time management, source definition, detector electronics modelling, and data output.

The {\it application layer} is composed of classes derived from the base classes of the core layer to model specific objects or processes. For example, the core layer defines the base class for volumes, and the application layer comprises all the derived classes for modeling specific volumes, including boxes, spheres, cylinders, and trapezoids. Similarly, the application layer includes all the specific movement models derived from the movement base class, including translations, rotations, orbits, and oscillations. Thus, the range of features available in \gate~can increase as new application classes are developed, while the general structure remains unaffected.

\begin{figure}[h]
\begin{center}
\includegraphics[scale=0.6,angle=0]{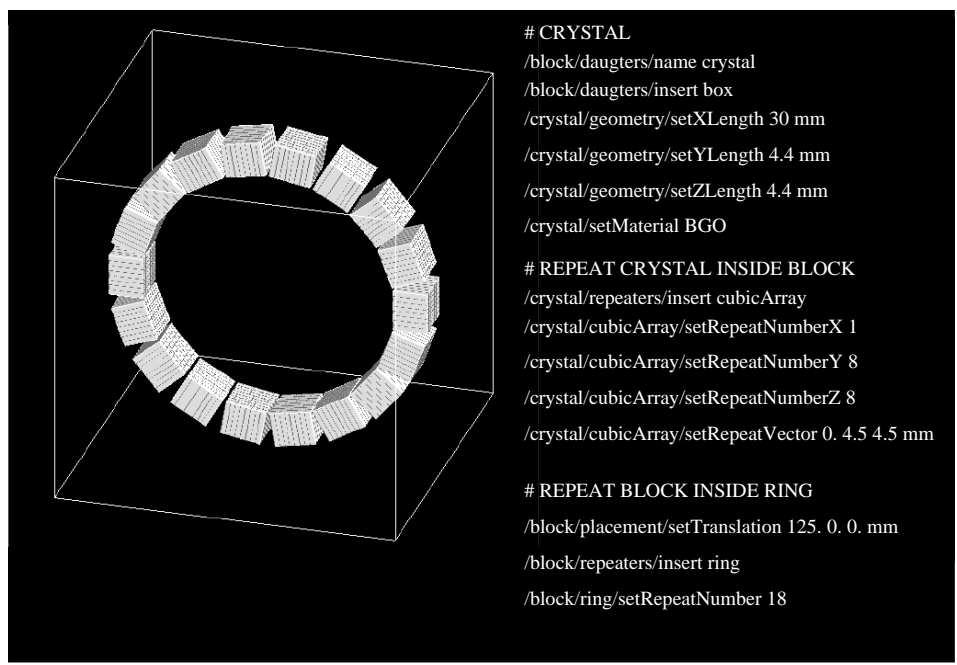}
\caption{Ring detector geometry obtained by using the command scripts displayed in the figure. This script models a detector-block, where $4.4\times4.4$~mm BGO crystals of 30~mm thickness are repeated in an $8\times8$~matrix. A {\it cubic array} repeater command produces an $8\times8$~cubic array of crystals from a single crystal with a crystal picth of 4.5~mm in both directions. A {\it ring} repeater places the block at 125.0 mm from the origin on the X-axis and replicates it eighteen times about a cylinder around the Z-axis.}
\label{fig:ring-repeaters}
\end{center}
\end{figure}

In the {\it user layer}, \geant~provides mechanisms for running simulations both interactively or batch-wise using scripts. An important principle of \gate~is that each class must provide dedicated extensions to the command interpreter class, so that the functionality provided by the class is available through script commands. The end-users of \gate~therefore do not have to perform any C++ coding. The complete set-up of a nuclear medicine experiment can be easily defined using the script language, as shown in Figure \ref{fig:ring-repeaters}.

\subsection{Systems}
\label{sec:sys}
When defining the geometry for a tomograph, specific guidelines with respect to the geometrical hierarchy of the tomograph components must be followed, so that the \geant~ particles interaction histories, called {\it hits}, occuring in the detector, can be processed to realistically mimic detector output. Most PET scanners are built following comparable concepts: one or more rings, each ring consisting of several scintillator blocks, each block being subdivided in crystal pixels. For SPECT, similar concepts exist: a gamma camera with a continuous or pixelated crystal, and a collimator. Most of these geometrical concepts are common to many different imaging systems. To facilitate the hierarchical description of a tomograph, predefined global {\it systems} are used. A {\it system} is defined as a family of geometries compatible with one or several data output formats. The main property of a {\it system} is that its geometry description is supported by specific list-mode or histogrammed (sinogram or projection) data output formats. Currently there are five {\it systems} available in \gate: one for SPECT, three for PET $-$ two for block detector geometries and one for continuous pixellated geometries $-$ and a generic system appropriate to model novel tomographic paradigms. This latter system is completely open and provides only basic building blocks for the definition of a tomographic experiment.

\subsection{Management of time and movements}
\label{sec:time}
\label{sec:movements}
\label{subsubsec:sdmov}
%
%

One of the distinctive features of \gate~is the management of time-dependent phenomena (Santin \etal 2003, Strul \etal 2003). The synchronization of the source kinetics with the movement of the geometry thus allows for the simulation of realistic acquisition conditions including patient movement, respiratory and cardiac motions, scanner rotation, or changes in activity distribution over time. Dealing with time in \gate~includes: (a) defining the movements associated with the physical volumes that describe the detector and phantom; (b) describing the radioactive sources; and (c) specifying the start and stop times of the acquisition (which are equivalent to the start and stop times in a real experiment).

The \geant~geometry architecture requires the geometry to be static during a simulation. However, the typical duration of a single event is very short when compared to movements in the geometry model or bio-kinetics. Movements are synchronized with the evolution of the source activities by subdividing the acquisition time frames (typically of the order of minutes or hours) into smaller time steps. At the beginning of each time step, the geometry is updated according to the requested movements. During each time step, the geometry is held at rest and the simulation of the particle transport proceeds. Within the time steps, the source is allowed to decay so that the number of events decreases exponentially from one time step to the next, and within the time steps themselves. The proper timing of the simulated event sequence is a key feature for modeling time-dependent processes such as count rates, random coincidences, event pile-up, and detector deadtime (Simon \etal 2004). Between time steps, the position and the orientation of a subset of daughter volumes can be changed to mimic a movement such as a rotation or a translation. These displacements are parametrized by their velocity. It is the responsibility of the user to set the time step duration short enough to produce smooth changes. Combinations of translations and rotations allow the simulation of complex acquisition trajectories of the detectors such as parameterized eccentric rotations. 

\geant~does not allow the movement of sources. Therefore, in \gate, an emission volume is defined so that it encompasses the actual source's volume throughout its range of displacement. To enable movement of the activity distribution, an additional volume is defined to confine the emission within the intersection of the emission and the confinement volumes. This confinement volume defines the shape and size of the actual source and moves within the emission volume.

%% file: pmb.sec.6.tex
\section{Physics}
\label{sec:phy}

\label{sec:geo}
%
%

\subsection{Radioactive sources}
\label{sec:radioactivesources}
A source in \gate~is defined by its particle type (\eg~radionuclide, gamma, positrons, etc.), position (volume), direction (solid angle), energy (spectrum), and activity. The lifetime of a radioactive source is usually obtained from the \geant~database, but it can also be set by the user to approximate a decay source through the emission of its decay products (\eg~positrons or gammas).

The activity determines the decay rate for a given source during the simulated acquisition time. Radioactive decay of radionuclides with secondary particle emission is performed by the \geant~Radioactive Decay Module (RDM), which has been modified so that \gate~source manager maintains control over the definition of decay time. Continuous event time flow is obtained by using a virtual clock that defines an absolute time $t$ used to initialize the \geant~internal tracking time. Random time intervals $\delta t$ between events that occur at time $t$ are sampled from a exponential distribution:
\begin{equation}
\label{eq:poisson}
p (\delta t) = A(t)~\exp ( - A(t) ~ \delta t)
\end{equation}
where $A(t)=A_0 \exp((t-t_0)~/~\tau)$ is the source activity at time $t$, $A_0$ is the user defined source activity at time $t_0$ and $\tau$ the lifetime. When the resulting decay time exceeds the end of the current time step, the run is terminated and a new one is started, allowing for the synchronization of the sources with the geometry movements. Multiple sources can be defined with independent properties. For each event, a proposed time interval is sampled for each source according to Eq.~\ref{eq:poisson}, and the shortest one is chosen for the primary generation. The overall behavior of this mechanism is such that, at all times, the relative importance of each source is proportional to its activity, while the overall time interval sampling is determined by the total activity of all sources. Voxelized phantom or patient data can be used as sources to reproduce realistic acquisitions: emission data are converted into activity levels, and \gate~can read in voxelized attenuation map and converts the gray scale into material definitions using an analogous translator.

 
\subsection{Positron emission}
\gate~ includes 2 modules dedicated to PET (Jan 2002). The first uses the von Neumann algorithm (von Neumann 1951) to randomly generate the positron energy according to the measured $\beta^{+}$ spectra. This method greatly increases the speed of the simulation by bypassing the decay of radionuclides process used by \geant. 
The $\beta^{+}$ spectra of 3 commonly used radionuclides ($^{11}$C,$^{15}$O, and $^{18}$F) have been parametrized in \gate~according to the Landolt-B\"ornstein tables~(Behrens and J\"anecke 1969). 

The second module deals with the acollinearity of the two annihilation photons, which is not accounted for in \geant~. In \gate, acollinearity is modelled using a 0.58$^{o}$ full width at half maximum (FWHM) Gaussian blur. This width corresponds to experimental values measured in water~(Iwata \etal 1997).  
%
%
\subsection{Interaction modelling with standard energy and low energy packages}
A material database file contains all material parameters required by \geant~to calculate the interaction cross-sections and is easily modified by the user. In contrast to \geant, \gate~only uses natural isotopic abundances. The fact that these cannot be modified by the user has little bearing on \gate~applications since isotopic abundances are unimportant in low to mid-energy photon and charged particle interactions.
The electromagnetic interactions used in \gate~are derived from \geant~. The electromagnetic physics package manages electrons, positrons, $\gamma$-rays, X-rays, optical photons, muons, hadrons, and nuclei. As in \geant, \gate~can use two different packages to simulate electromagnetic processes: the standard energy package, and the low energy package. In the standard energy package, photoelectric effect and Compton scatter can be simulated at energies above 10~keV. Under 100~keV however, relative errors on the cross-sections are higher than 10\% and can raise up above 50\% (Jan 2002 and Lazaro 2003). The low energy package models photon and electron interactions down to 250 eV and includes Rayleigh scattering. For biomedical applications, it provides more accurate models for the electromagnetic interactions. However, this comes at the price of increased computing time.

\subsection {Secondary production cuts}

\gate~inherits the \geant~capability to set thresholds for the production of secondary electrons, X-rays and delta-rays (Agnostelli \etal 2003). In biomedical applications, eliminating the secondary particles whose initial energy is below the production threshold increases the computing efficiency.

Because low energy processes generate more secondary particles than standard energy processes, cuts affect simulation speed more strongly when applied with the low energy package. Turning off the production of electrons, X-rays, and delta-rays by setting high thresholds may result in a substantial increase in computing speed for a typical simulation of a PET scanner. In many cases, the accuracy of the simulation at the level of single or coincidence photon counting is preserved. 

%% file: pmb.sec.7.tex
\section{Digitization}
\label{sec:digi}
%
%
%

Digitization is the process of simulating the electronics response of a detector within a scanner. This involves the conversion of the charged particle and photon interactions into energy bins, detection positions, and coincidences. To do this, portions of the scanner geometry are designated as {\it sensitive detectors}, which record interactions within these regions. The {\it digitizer chain} then processes these recorded interactions and produces counts and coincidences. The sensitive detectors and digitizer chain are described below.

\subsection{Sensitive detectors}
\label{sub_sec:sentitive_detectors}

\begin{figure}
\begin{center}
\includegraphics[scale=0.5,angle=0]{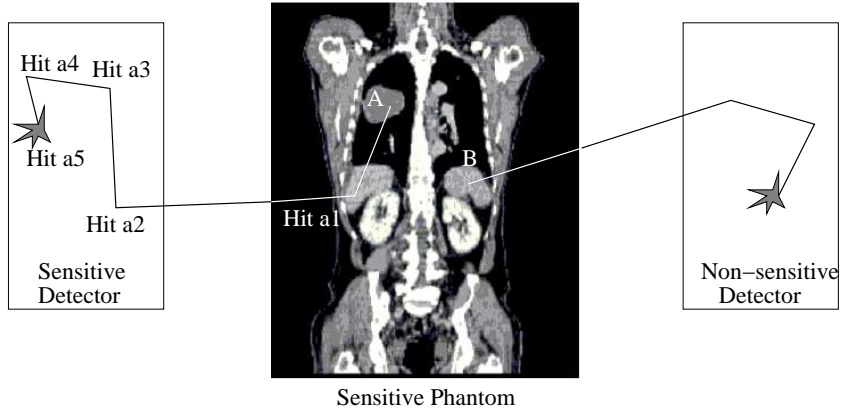}
\caption{Particle interactions in a {\it crystalSD} attached to a scintillator block and a {\it phantomSD} attached to a volume filled with tissue material. The trajectory of particle A shows 1 hit in the {\it phantomSD} (Hit~a1) and 4 hits in the {\it crystalSD} (Hit~a2 to Hit~a5). Particle B does not interact within a sensitive volume, thus no hit information is stored.}
\label{fig:SD}
\end{center}
\end{figure}

Sensitive detectors are used to store information about particle interactions (hereafter referred to as {\it hits}) within volumes. \gate~only stores hits for those volumes that have a sensitive detector attached. Two types of sensitive detectors are defined in \gate: the crystal sensitive detector ({\it crystalSD}) is used to generate hits from interactions that occur inside the detector portions of the scanner (Figure~\ref{fig:SD}). The phantom  sensitive detector ({\it phantomSD}) is used to detect and count the Compton and Rayleigh interactions occurring within the scanner's field-of-view (FOV).

\subsection{Digitizer chain}
The digitizer chain mimics a realistic detection process by building the physical observables from the hits. The observables of each detected event are energy, position, and time of the interaction. The digitizer consists of a chain of processing modules that takes a list of hits from the sensitive detectors and transforms them into pulses referred to as {\it singles}. The key elements of this chain are now briefly described.

\subsubsection{Hit adder} A particle entering a sensitive detector can generate multiple hits, as shown in Figure~\ref{fig:SD}. For instance, a gamma ray interacting within a scintillation crystal can generate hits corresponding to several Compton scattering events and a photoelectric absorption. The {\it hit adder} sums the deposited energy of these hits within a sensitive detector to yield a {\it pulse}. The position of the pulse is calculated from the energy-weighted centroid of the hit positions, and the time of the pulse is set to that of the first hit within the volume. If a particle interacts in several sensitive detectors the hit adder will generate a list of pulses, one for each sensitive detector.

\subsubsection{Pulse reader} With the exception of one-to-one readout schemes, scanners often have a readout segmentation different from the detector segmentation. To simulate this, a {\it pulse reader} module adds the pulses together within a user-defined group of sensitive detectors. This yields a pulse containing the total energy deposited within the group of sensitive detectors. The position of this pulse is set to that of the pulse from the adder that has the largest energy (winner-takes-all paradigm). 

\subsubsection{User modules}  Following the hit adder and the pulse reader, which regroup the hits into pulses and then sum pulses, the remaining modules of the digitizer chain transform these pulses into the physical observables of the scanner (\ie~singles). These modules are discussed below.

\begin{description}
\item[Energy resolution:]  This module applies a Gaussian blur to the energy of the pulse, $E$, with an FWHM of ${R \times E}$. The FWHM energy resolution, $R$, is given by:
\begin{equation}
\label{eq:resolution}
R=\frac{R_0\sqrt{E_0}}{\sqrt{E}}
\end{equation}
where $R_0$ is the user defined FWHM energy resolution at a given energy, $E_0$.\\

A more elaborate model propagates the relative variances of the physical processes involved with light collection and detection in a spectrometric chain as:
\begin{equation}
\label{eq:propagation}
{\upsilon_E}={\upsilon_N}+\frac{\upsilon_\eta}{\bar{N}}+\frac{\upsilon_\epsilon}{\bar{N} \eta}+\frac{\upsilon_M}{\bar{N} \eta \epsilon}
\end{equation}
where $\upsilon_E$, $\upsilon_N$, $\upsilon_\eta$, $\upsilon_\epsilon$, and $\upsilon_M$ are the relative variances on $E$, on the number of scintillation photons $N$, on the light collection efficiency $\eta$, on the quantum efficiency of the photo-detector $\epsilon$, and on the gain of the photo-detector $M$. Both the light collection efficiency and the quantum efficiency are binomial processes with probabilities of $\eta$ and $\epsilon$, respectively. If the scintillation process is assumed to follow a Poisson law with a mean equal to ${\bar{N} = E \times LY}$, where $LY$ is the light yield of the scintillator, then relative variance on $E$ is given by:
\begin{equation}
\label{eq:relative_variance}
{\upsilon_E}=\frac{1 + \upsilon_M}{\bar{N} \eta \epsilon}
\end{equation}
In the case of a photomultiplier, $\upsilon_M$ is approximately 0.1 and the FWHM energy resolution is estimated using the equation:
\begin{equation}
\label{eq:resolution_bis}
R=\sqrt{{2.35}^2 \frac{1.1}{\bar{N} \eta \epsilon} + {R_i}^2}
\end{equation}
where $R_i$ is the intrinsic resolution of the scintillator (Kuntner \etal 2002).

\item[Energy window:] Upper and lower energy thresholds can be set for several energy windows by using multiple processor chains. These thresholds are applied using either a step or sigmoid function.

\item[Spatial resolution:] For SPECT, spatial resolution is modeled using a Gaussian blur of the position. For PET, interaction position is calculated by the pulse reader which simulates the intrinsic spatial resolution of the detector. More elaborate models can be derived for continuous crystal detector PET systems (Staelens \etal 2004). These models are currently under development.

\item[Time resolution:] Simulation of time jitter can be obtained using a Gaussian blur of the pulse time.

\item[Deadtime:] Both paralyzable and non-paralyzable deadtimes can be modeled explicitely on an event-by-event basis. While these models represent the idealized behavior, they correctly predict the theoretical lifetimes for both types of deadtimes (Simon \etal~2004).

\item[Coincidence:] At the end of a digitizer chain a coincidence sort can be added to find pairs of singles that are in coincidence. Pairs of singles can be considered coincidences whenever the time interval between the singles is less than a user-defined coincidence window. Each single is stored with its corresponding event number. If the event numbers of the singles associated in a coincidence are different, this is a random coincidence. A similar flag exists for Compton scattered events. The Compton scatter flag can be used to differentiate true from scattered coincidence pairs that have identical event flags. Multiple coincidences corresponding to more than 2 singles within the same coincidence window are discarded.
\end{description}

%
%

\subsection{\dig}
\label{debug}
In \gate~standard operation mode, primary particles are generated by a source manager and then propagated through the attenuating geometry before generating hits in the sensitive detectors, which are then processed by the digitizer chain. While this operating mode is suitable for many purposes, it is inefficient for the optimization of the digitizer chain parameters. This is best done by comparing the results from different sets of digitizer parameters using the same series of hits. To perform this specific task, \gate~offers an operating mode named {\it \dig}. In this mode, hits are read from a data file generated by \gate~and fed directly into the digitizer chain. The same command scripts are used for both the hit generation simulation and {\it \dig}~simulations. Thus, all conditions are kept identical in the simulations including time-dependence.

%% file: pmb.sec.9.tex
\section{Simulation benchmarks}
\label{sec:simbench}
%
%

Two benchmarks, one for PET and one for SPECT, are included in the \gate~distribution. These benchmarks check the integrity of the installation or upgrade, and also allow for the comparison of CPU performance on different computing platforms. In addition, they provide examples of how to use the main features of \gate~to simulate PET or SPECT experiments. Furthermore, they serve as examples on how to analyze output. Each benchmark consists of macros to run the simulation, analyze simulation output, and generate figures. In addition, a set of baseline figures are included for a comparison between the user's results with those from a correct run.

\subsection{PET benchmark}
\label{subsec:PETI}

\begin{figure}
\begin{center}
\includegraphics[width=.60\textwidth]{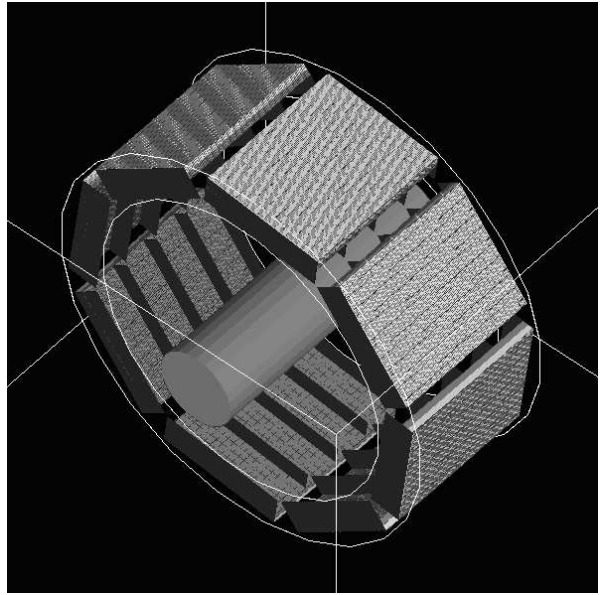}
\caption{\label{fig:petbench} Illustration of the PET benchmark setup.} 
\end{center}
\end{figure}

The PET benchmark (Figure~\ref{fig:petbench}) simulates a whole-body scanner that does not correspond to any existing system. Rather, it serves as a simple system that contains the majority of \gate~features. It consists of eight detector heads arranged within a 88~cm diameter by 40~cm axial length octagonal cylinder. Each head is made of 400 detector blocks and each block is a $5 \times 5$ array of dual-layer LSO-BGO crystals. The heads are equipped with partial septa that rotate in a step-and-shoot mode. The phantom in this benchmark is a 70~cm long water cylinder with one $^{18}$F (half-life~=~109.8~min) and one $^{15}$O (half-life~=~2.03 min) line source each with an activity of 100~kBq. The simulated acquisition is 4 min in duration, which represents approximately two $^{15}$O half-lives. The source activities are set such that the PET benchmark will run in about 12~CPU hours on a 1~GHz processor. The acquisition is divided into two 2 min frames; after the first frame, the gantry rotates by 22.5 deg. Only coincident events are recorded, using a coincidence time window of 120~ns. This large window is used in order to record a large number of random coincidences. The lower and upper energy thresholds are set to 350 and 650~keV, respectively.

The standard electromagnetic package of \geant~is used in this benchmark. Only the Rayleigh interactions are modeled using the low energy package. To speed up the simulation, X-rays and secondary electrons are not tracked.

Approximately $3.7\times 10^7$ decays occur during the simulated acquisition and around $7.0\times 10^5$ coincidences are recorded and written in a \rt~file (Brun and Rademakers 1997). Based on the \rt~output, several figures and plots are calculated using \rt~ to confirm the correct execution of the simulation. The benchmark results are characterized by: (1) the total number of generated events and detected coincidences; (2) their spatial and time distributions; (3) the fractions of random and scattered coincidences; and (4) the average acollinearity between the two annihilation gammas.

The PET benchmark has been run on 12 different system configurations. Two operating systems were tested : Linux and Mac OS (versions 10.2.6 and 10.3). The Linux distributions were RedHat (versions 9.0, 8.0 and 7.3) and SuSE (version 8.1). The source code compilation was performed with either gcc 2.95 or 3.2. The computing time for the PET benchmark averaged around 12 and 6 hours for 1.0 GHz and 3 GHz processors, respectively.

PET physical variables characterizing the simulation results are shown in Table~\ref{PETbench} with their mean value and standard deviation obtained from the run of the PET benchmark on the 12 system configurations, using a different initial seed on each system for the generation of random numbers.
\begin{table}[htb!]
\caption{Average values and relative standard deviations (stdev) of the physical variables studied with the PET benchmark.}
\label{PETbench}
\begin{center}
\begin{tabular}{lrr}
\hline \hline
\multicolumn{1}{l}{\footnotesize  variable type} & \multicolumn{1}{r}{\footnotesize  average value} &
\multicolumn{1}{r}{\footnotesize  relative stdev} \\
\hline
\multicolumn{1}{l}{\footnotesize total decays during the acquisition} & 
\multicolumn{1}{r}{\footnotesize  $3.6815\times 10^7$} & \multicolumn{1}{r}{\footnotesize $\pm 0.01\%$} \\
\multicolumn{1}{l}{\footnotesize random coincidences} & 
\multicolumn{1}{r}{\footnotesize $20,568$} & \multicolumn{1}{r}{\footnotesize $\pm 0.58\%$} \\ 
\multicolumn{1}{l}{\footnotesize unscattered coincidences} & 
\multicolumn{1}{r}{\footnotesize $311,778$}  & \multicolumn{1}{r}{\footnotesize $\pm 0.39\%$} \\ 
\multicolumn{1}{l}{\footnotesize scattered coincidences} & 
\multicolumn{1}{r}{\footnotesize $369,915$} & \multicolumn{1}{r}{\footnotesize $\pm 0.33\%$} \\ 
\multicolumn{1}{l}{\footnotesize simulated 15-Oxygen life time} & 
\multicolumn{1}{r}{\footnotesize $123.316$} s & \multicolumn{1}{r}{\footnotesize $\pm 0.06\%$} \\ 
\multicolumn{1}{l}{\footnotesize gamma acollinearity angle} & 
\multicolumn{1}{r}{\footnotesize $0.6063$ deg} & \multicolumn{1}{r}{\footnotesize $\pm 0.01\%$} \\
\hline \hline
\end{tabular}
\end{center}
\end{table}
The results in Table~\ref{PETbench} and, in particular, the relative standard deviations show that the main physical simulation variables are stable within less than 1\%. It is strongly recommended that the user validates a new or updated \gate~installation using this table.

\subsection{SPECT benchmark}
\label{subsec:SPECTI}
\begin{figure}
\begin{center}
\includegraphics[width=.60\textwidth]{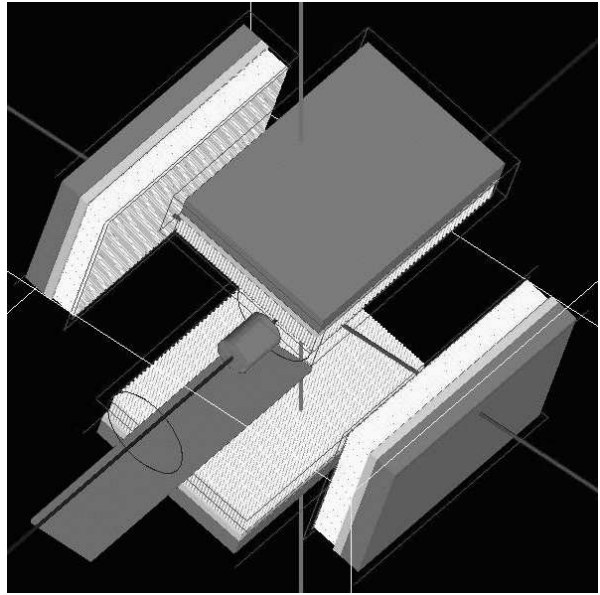}
\caption{\label{fig:spectbench} Illustration of the SPECT benchmark setup.} 
\end{center}
\end{figure}

The SPECT benchmark (Figure~\ref{fig:spectbench}) simulates a SPECT acquisition with a moving source. The simulated 4-head gamma camera does not correspond to any real system. This benchmark involves a cylindrical phantom (5~cm in diameter and 20~cm long) filled with water with an inner cylinder (2~cm in diameter, 5~cm long) filled with 30~kBq of $^{99m}$Tc. The phantom lies on a table (0.6~cm thick, 8~cm wide, and 34~cm long). During the simulated acquisition, the table and phantom translate together at 0.04~cm/s. Confinement is used to keep the source distribution synchronized with the phantom movement. All 4 heads of the gamma camera are identical, consisting of a parallel hole lead collimator (hole diameter: 0.3~cm, collimator thickness: 3~cm, and septa thickness: 0.6~mm), a 1~cm thick NaI crystal, a 2.5~cm thick back-compartment in Perspex, and a 2~cm thick lead shielding. The simulated acquisition consists of 64 projections (16 projections per head), acquired along a circular orbit with a 7~cm radius of rotation and a speed of 0.15~deg/s. Sixteen runs of 37.5~s each are performed to simulate the 16 positions of the 4 gamma camera heads.\\
The low energy electromagnetic processes package is used to model the physics. Rayleigh, photoelectric, and Compton interactions are turned on while the gamma conversion interactions are turned off. To speed up the simulation, the X-ray production cut is set at 20 keV and secondary electrons are not tracked. Compton events occurring in the phantom, collimator, back-compartment, shielding and table are recorded. A Gaussian energy blur is applied to all events detected in the crystal, using an energy resolution of 10\% at 140~keV. The limited spatial resolution of the photomultipliers and associated electronics is modeled using a Gaussian blur with a standard deviation of 2~mm. Only photons detected with an energy between 20 and 190 keV are stored.

The benchmark results are characterized by: (1) the number of simulated events and of detected counts; (2) the percentage of primary and scattered events with respect to all events detected in the 20-190 keV energy window (here scattered events are considered a function of the compartment in which the last scattered event occurred \ie~phantom, collimator, table, crystal, or back-compartment); (3) the mean and standard deviation of the number of detected counts per projection; (4) the percentage of scattered events as a function of the scattering order ($1^{st}$ for single scatter, $2^{nd}$ for double scatter, and so on).

\begin{table}[htb!]
\caption{Average values and relative standard deviations (stdev) of the figures of merit used for the SPECT benchmark.}
\label{SPECTbench}
\begin{center}
\begin{tabular}{lrr}
\hline \hline
\multicolumn{3}{l}{\bf\footnotesize Global information} \\
\multicolumn{1}{l}{\footnotesize  variable type} & \multicolumn{1}{r}{\footnotesize  average value} & \multicolumn{1}{r}{\footnotesize  relative stdev} \\
\hline
\multicolumn{1}{l}{\footnotesize  number of emitted particles} & \multicolumn{1}{r}{\footnotesize  $1.79994 \times 10^7$} & \multicolumn{1}{r}{\footnotesize  $\pm 0.01\%$} \\
\multicolumn{1}{l}{\footnotesize  detected counts between 20 and 190 keV} & \multicolumn{1}{r}{\footnotesize  35,919} & \multicolumn{1}{r}{\footnotesize  $\pm 0.3\%$} \\
\multicolumn{1}{l}{\footnotesize  percentage of unscattered photons} & \multicolumn{1}{r}{\footnotesize  32.9\%} & \multicolumn{1}{r}{\footnotesize  $\pm 1.0\%$} \\
\multicolumn{1}{l}{\footnotesize  mean detected counts per projection} & \multicolumn{1}{r}{\footnotesize  278.4} & \multicolumn{1}{r}{\footnotesize  $\pm 0.9\%$} \\
\hline \hline
\multicolumn{3}{l}{\bf\footnotesize  Percentage of photons whose last scattered event occurred in a specific medium }\\
\multicolumn{1}{l}{\footnotesize  medium} & \multicolumn{1}{r}{\footnotesize  average value} & \multicolumn{1}{r}{\footnotesize  relative stdev} \\
\hline
\multicolumn{1}{l}{\footnotesize  phantom} & \multicolumn{1}{r}{\footnotesize  52.3\%} & \multicolumn{1}{r}{\footnotesize  $\pm 0.6\%$} \\
\multicolumn{1}{l}{\footnotesize  table} & \multicolumn{1}{r}{\footnotesize  3.1\%} & \multicolumn{1}{r}{\footnotesize  $\pm 1.8\%$} \\
\multicolumn{1}{l}{\footnotesize  collimator} & \multicolumn{1}{r}{\footnotesize  2.1\%} & \multicolumn{1}{r}{\footnotesize  $\pm 2.4\%$} \\
\multicolumn{1}{l}{\footnotesize  crystal} & \multicolumn{1}{r}{\footnotesize  8.5\%} & \multicolumn{1}{r}{\footnotesize  $\pm 1.2\%$} \\
\multicolumn{1}{l}{\footnotesize  back-compartment} & \multicolumn{1}{r}{\footnotesize  1.2\%} & \multicolumn{1}{r}{\footnotesize  $\pm 3.3\%$} \\
\hline \hline
\multicolumn{3}{l}{\bf\footnotesize  Percentage of scattered photons as a function of the scattering order }\\
\multicolumn{1}{l}{\footnotesize  scattering order} & \multicolumn{1}{r}{\footnotesize  average value} & \multicolumn{1}{r}{\footnotesize  relative stdev} \\
\hline
\multicolumn{1}{l}{\footnotesize  order 1} & \multicolumn{1}{r}{\footnotesize  46.4\%} & \multicolumn{1}{r}{\footnotesize  $\pm 0.7\%$} \\
\multicolumn{1}{l}{\footnotesize  order 2} & \multicolumn{1}{r}{\footnotesize  26.8\%} & \multicolumn{1}{r}{\footnotesize  $\pm 0.7\%$} \\
\multicolumn{1}{l}{\footnotesize  order 3} & \multicolumn{1}{r}{\footnotesize  13.6\%} & \multicolumn{1}{r}{\footnotesize  $\pm 1.5\%$} \\
\multicolumn{1}{l}{\footnotesize  order 4} & \multicolumn{1}{r}{\footnotesize  6.7\%} & \multicolumn{1}{r}{\footnotesize  $\pm 0.7\%$} \\
\multicolumn{1}{l}{\footnotesize  order $>$ 4} & \multicolumn{1}{r}{\footnotesize  6.5\%} & \multicolumn{1}{r}{\footnotesize  $\pm 1.6\%$} \\
\hline \hline
\end{tabular}
\end{center}
\end{table}
The SPECT benchmark has run on 11 system configurations. Two operating systems have been tested: Linux (RedHat versions 9.0, 8.0, 7.3, 7.1, SuSE versions 9.0 and 8.1, and Fedora Core 1) and Mac OS (version 10.2.8). The mean and standard deviation of most figures of merit characterizing the results of the runs are given in Table~\ref{SPECTbench}. Similar to the PET benchmark results, this table shows that the results produced by \gate~are very stable. Table~\ref{SPECTbench} should be used to validate any new installation or update of \gate. The time needed to run the benchmark ranged from about 3 hours (2.8 GHz Pentium, 2 Gb RAM) to about 11 hours (1 GHz Pentium, 2.3 Gb RAM).

%% file: pmb.sec.10.tex
\section{Validation of \gate}
\label{sec:valGATE}
%
%

The validation of Monte Carlo simulated data against real data obtained with PET and SPECT cameras is essential to assess the accuracy of \gate~ and the \OG~collaboration is largely involved into the validation of \gate. Tables \ref{validationPET} and \ref{validationSPECT} list commercial systems which have been or are currently being considered for PET and SPECT validations. These tables summarize the figures of merit (FOM) used for assessing the consistency between simulated and real data, as well as the major validation results and associated references. For details regarding these validation studies, the reader is highly encouraged to refer to the appropriate references. Overall, these studies illustrate the flexibility and reliability of \gate~for accurately modelling various detector designs. The modelling of the Millennium VG (GE MS) is also in progress.

\begin{table}[htb!]
\caption{Validation result summary of commercial systems already or currently considered for \gate~validation in PET.}
\label{validationPET}
\begin{center}
\begin{tabular}{llll}
\hline\hline
\multicolumn{1}{l}{\footnotesize  PET system} & \multicolumn{1}{l}{\footnotesize Studied FOM} & 
\multicolumn{1}{l}{\footnotesize Experiment/GATE} & \multicolumn{1}{l}{\footnotesize References}\\ 
\hline
\multicolumn{1}{l}{\footnotesize ECAT EXACT HR$^{+}$,} & \multicolumn{1}{l}{\footnotesize Spatial resolution} & & \multicolumn{1}{l}{\footnotesize Jan \etal 2003}\\
\multicolumn{1}{l}{\footnotesize CPS} & \multicolumn{1}{l}{\footnotesize    - radial @10cm} & \multicolumn{1}{l}{\footnotesize 11.4 mm / 11.8 mm} & \\
& \multicolumn{1}{l}{\footnotesize    - tangential @10cm} & \multicolumn{1}{l}{\footnotesize 11.8 mm / 10.7 mm} & \\
& \multicolumn{1}{l}{\footnotesize 3D sensitivity} & \multicolumn{1}{l}{\footnotesize 0.75\% / 0.80\%} & \\
& \multicolumn{1}{l}{\footnotesize 3D count rates} & & \\
& \multicolumn{1}{l}{\footnotesize    - prompts @10kBq/ml} & \multicolumn{1}{l}{\footnotesize 550 kcps / 550 kcps} & \\
& \multicolumn{1}{l}{\footnotesize    - trues @10kBq/ml} & \multicolumn{1}{l}{\footnotesize 330 kcps / 300 kcps} & \\
& \multicolumn{1}{l}{\footnotesize 3D scatter fraction} & \multicolumn{1}{l}{\footnotesize 36\% / 35\%} & \\
\hline
\multicolumn{1}{l}{\footnotesize Allegro, Philips} & \multicolumn{1}{l}{\footnotesize 3D count rate} & \multicolumn{1}{l}{\footnotesize @555MBq/ml} & \multicolumn{1}{l}{\footnotesize Lamare \etal}\\
& \multicolumn{1}{l}{\footnotesize    - trues} & \multicolumn{1}{l}{\footnotesize 800 kcps / 950 kcps} & \multicolumn{1}{l}{\footnotesize 2004}\\
& \multicolumn{1}{l}{\footnotesize    - scatter} & \multicolumn{1}{l}{\footnotesize 950 kcps / 900 kcps} & \\
& \multicolumn{1}{l}{\footnotesize    - randoms} & \multicolumn{1}{l}{\footnotesize 2,000 kcps / 2,400 kcps} & \\
& \multicolumn{1}{l}{\footnotesize 3D scatter fraction} & \multicolumn{1}{l}{\footnotesize 8\% difference} & \\
\hline 
\multicolumn{1}{l}{\footnotesize GE Advance, GE MS} & \multicolumn{1}{l}{\footnotesize Energy spectra} &
\multicolumn{1}{l}{\footnotesize visual assessment} & \multicolumn{1}{l}{\footnotesize Schmidtlein \etal}\\
& \multicolumn{1}{l}{\footnotesize 3D scatter fraction} & \multicolumn{1}{l}{\footnotesize 47.1\% / 47.2\%} & \multicolumn{1}{l}{\footnotesize 2004}\\
\hline 
\multicolumn{1}{l}{\footnotesize MicroPET 4,} & \multicolumn{1}{l}{\footnotesize Spatial resolution} & & \multicolumn{1}{l}{\footnotesize Jan \etal 2003b}\\
\multicolumn{1}{l}{\footnotesize Concorde} & \multicolumn{1}{l}{\footnotesize    - radial @2cm} & \multicolumn{1}{l}{\footnotesize 2.35 mm / 2.25 mm} & \\
& \multicolumn{1}{l}{\footnotesize    - tangential @2cm} & \multicolumn{1}{l}{\footnotesize 2.45 mm / 2.30 mm} & \\
& \multicolumn{1}{l}{\footnotesize Sensitivity (350-650 keV)} & \multicolumn{1}{l}{\footnotesize 1.43\% / 2.42\%} & \\
& \multicolumn{1}{l}{\footnotesize Miniature Derenzo} & \multicolumn{1}{l}{\footnotesize visual assessment} & \\
\hline 
\multicolumn{1}{l}{\footnotesize MicroPET Focus,} & \multicolumn{1}{l}{\footnotesize Spatial resolution} & & \multicolumn{1}{l}{\footnotesize Jan \etal 2004}\\
\multicolumn{1}{l}{\footnotesize Concorde} & \multicolumn{1}{l}{\footnotesize    - radial @8cm} & \multicolumn{1}{l}{\footnotesize 3.55 mm / 3.4 mm} & \\
& \multicolumn{1}{l}{\footnotesize Sensitivity} & \multicolumn{1}{l}{\footnotesize 3.4\% / 3.5\%} & \\
\hline
\end{tabular}
\end{center}
\end{table}

\begin{table}[htb!]
\caption{Validation result summary of commercial systems already or currently considered for \gate~validation in SPECT.}
\label{validationSPECT}
\begin{center}
\begin{tabular}{llll}
\hline\hline
\multicolumn{1}{l}{\footnotesize SPECT system} & \multicolumn{1}{l}{\footnotesize Studied FOM} & 
\multicolumn{1}{l}{\footnotesize Experiment/GATE} & \multicolumn{1}{l}{\footnotesize References}\\
\hline
\multicolumn{1}{l}{\footnotesize IRIX, Philips} & \multicolumn{1}{l}{\footnotesize None reported} & 
\multicolumn{1}{l}{\footnotesize n/a} & \multicolumn{1}{l}{\footnotesize Staelens \etal 2004b}\\
\hline
\multicolumn{1}{l}{\footnotesize AXIS, Philips} & \multicolumn{1}{l}{\footnotesize Spatial resolution} &
\multicolumn{1}{l}{\footnotesize 1.30 cm / 1.36 cm} & \multicolumn{1}{l}{\footnotesize Staelens \etal 2003}\\
& \multicolumn{1}{l}{\footnotesize Energy spectra} & \multicolumn{1}{l}{\footnotesize visual assessment} & \\
& \multicolumn{1}{l}{\footnotesize Sensitivity} & \multicolumn{1}{l}{\footnotesize 231 cps/MBq / 246 cps/MBq} & \\
& \multicolumn{1}{l}{\footnotesize Scatter profiles} & \multicolumn{1}{l}{\footnotesize visual assessment} & \\
\hline
\multicolumn{1}{l}{\footnotesize Solstice, Philips} & \multicolumn{1}{l}{\footnotesize Sensitivity} & 
\multicolumn{1}{l}{\footnotesize good agreement} & \multicolumn{1}{l}{\footnotesize Staelens \etal 2003}\\
&  & \multicolumn{1}{l}{\footnotesize with theoretical data} & \multicolumn{1}{l}{\footnotesize Staelens \etal 2004c} \\
& & & \multicolumn{1}{l}{\footnotesize Staelens \etal 2004d}\\
\hline 
\multicolumn{1}{l}{\footnotesize DST Xli, GEMS} & \multicolumn{1}{l}{\footnotesize Energy spectra} & 
\multicolumn{1}{l}{\footnotesize excellent agreement} & \multicolumn{1}{l}{\footnotesize Assi\'e \etal 2004b}\\
& \multicolumn{1}{l}{\footnotesize Spatial resolution} & & \\
& \multicolumn{1}{l}{\footnotesize    - @10cm in air} & \multicolumn{1}{l}{\footnotesize 9.5 mm / 9.6 mm}& \\
& \multicolumn{1}{l}{\footnotesize    - @20cm in water} & \multicolumn{1}{l}{\footnotesize 14.2 mm / 14.4 mm}& \\
& \multicolumn{1}{l}{\footnotesize Sensitivity} & \multicolumn{1}{l}{\footnotesize $<$4\% difference} & \\
\hline
\end{tabular}
\end{center}
\end{table}

\gate~has also been shown to be appropriate for simulating various prototype imaging devices dedicated to small animal imaging. Table \ref{validationproto} shows the prototypes currently simulated using \gate~and indicates the features that have been studied and validated against experimental measurements. 

\begin{table}[htb!]
\caption{Prototypes dedicated to small animal imaging modeled using \gate~and features that have been studied using simulated data and summary validation results when available.}
\label{validationproto}
\begin{center}
\begin{tabular}{llll}
\hline\hline
\multicolumn{1}{l}{\footnotesize Prototype} & \multicolumn{1}{l}{\footnotesize Studied FOM} &
\multicolumn{1}{l}{\footnotesize Experiment/GATE} & \multicolumn{1}{l}{\footnotesize Reference}\\
\hline
\multicolumn{1}{l}{\footnotesize LSO/LuYAP phoswich PET} & \multicolumn{1}{l}{\footnotesize Sensitivity} & 
\multicolumn{1}{l}{\footnotesize n/a} & \multicolumn{1}{l}{\footnotesize Rey \etal 2003}\\
& \multicolumn{1}{l}{\footnotesize NEC curves} & \multicolumn{1}{l}{\footnotesize n/a} & \\
\hline
\multicolumn{1}{l}{\footnotesize High resolution dual head PET} & \multicolumn{1}{l}{\footnotesize Spatial resolution} &  & \multicolumn{1}{l}{\footnotesize Chung \etal 2003} \\
& \multicolumn{1}{l}{\footnotesize    - at center} & \multicolumn{1}{l}{\footnotesize 1.60 mm / 1.55 mm}& \\
& \multicolumn{1}{l}{\footnotesize    - 4 mm off-center} & \multicolumn{1}{l}{\footnotesize 1.72 mm / 1.7 2mm}& \\
& \multicolumn{1}{l}{\footnotesize Sensitivity} & \multicolumn{1}{l}{\footnotesize 0.13\% / 0.12\%} &\\
& \multicolumn{1}{l}{\footnotesize Line phantom} & \multicolumn{1}{l}{\footnotesize visual assessment} &\\
\hline
\multicolumn{1}{l}{\footnotesize CsI(Tl) SPECT camera} & \multicolumn{1}{l}{\footnotesize Energy spectra} &
\multicolumn{1}{l}{\footnotesize good agreement} & \multicolumn{1}{l}{\footnotesize Lazaro \etal 2004} \\
& \multicolumn{1}{l}{\footnotesize Spatial resolution} &  &\\
& \multicolumn{1}{l}{\footnotesize    - @10cm in air} & \multicolumn{1}{l}{\footnotesize 6.7 mm / 6.8 mm}& \\
& \multicolumn{1}{l}{\footnotesize Scatter fraction} & \multicolumn{1}{l}{\footnotesize 0.531\% / 0.527\%} &\\
& \multicolumn{1}{l}{\footnotesize Sensitivity} & \multicolumn{1}{l}{\footnotesize $<$2\% difference } &\\
& \multicolumn{1}{l}{\footnotesize Line phantom} & \multicolumn{1}{l}{\footnotesize visual assessment} &\\
\hline 
\multicolumn{1}{l}{\footnotesize OPET} & \multicolumn{1}{l}{\footnotesize Spatial resolution} &
\multicolumn{1}{l}{\footnotesize n/a} & \multicolumn{1}{l}{\footnotesize Rannou \etal 2003} \\
& \multicolumn{1}{l}{\footnotesize Sensitivity} & \multicolumn{1}{l}{\footnotesize n/a} &\\
\hline
\end{tabular}
\end{center}
\end{table}

%% file: pmb.sec.11.tex
\section{Work in progress and future developments}
\label{sec:ogcollab}
%
%

\subsection{Speed-up techniques}

Compared to simpler dedicated codes like SimSET (Harrison \etal 1993), the versatility of \gate~comes at the expense of relatively long computation times. To compensate for this, variance reduction tools are currently being developed for \gate. Another approach to improve the computing performance of \gate~is to distribute the simulations on multiple architectures. This is referred to as the {\it gridification} of \gate, and consists of subdividing simulations on geographically distributed processors in a Grid environment by parallelizing the random number generator. \gate~simulations use a very long period pseudo-random number generator developed from the algorithm of F. James (James 1990, Marsaglia and Zaman 1987). This random number generator can be subdivided into 900 million different non-overlapping sub-sequences, each containing approximately 10$^{30}$ numbers. These sub-sequences provide convenient starting points throughout the main sequence. Parallel simulations are produced by using the sub-sequences as independent streams by a sequence splitting method (Traore and Hill 2001). To demonstrate the potential benefits of {\it gridification}, a simulation was performed. Table~\ref{cputime} gives the total computing time in minutes of a simulation running on a single 1.5 GHz Pentium IV and the same simulation split between 10, 20, 50 and 100 processors on the European DataGrid testbed. In this experiment, the testbed consisted of 200 dual-processors with a mix of 750 MHz and 1 GHz Pentium III, and 1.4 GHz Pentium IV processors. {\it Gridification} using 20 jobs decreased the computing time by a factor of $8$ when compared to using a single Pentium IV 1.5 GHz processor. This example emphasizes the fact that computing time is not proportional to the number of jobs running in parallel, due in part to the launch time of the jobs, the modelling of the geometry of the simulation (which is independent of random number generation), and the time spent in queuing jobs. This study also proved that the results obtained via the Grid were equivalent to those generated on a single machine. In the future, each Unix platform on the DataGrid will be installed with \gate, allowing the use of more than 500 processors for a simulation. The development of a convenient tool to split, launch, and retrieve \gate~simulations on a Grid environment using a web interface is currently under development.

\begin{table}[htb!]
\begin{center}
\caption{Comparison of computing times between local and parallel jobs.}
\label{cputime}
\begin{tabular}{lrr}
\hline\hline
\multicolumn{1}{l}{\footnotesize Total computing time in minutes} &\\
\hline    
\multicolumn{1}{l}{\footnotesize 1 1.5 GHz Pentium IV} & \multicolumn{1}{r}{\footnotesize 159}\\
\multicolumn{1}{l}{\footnotesize 10 jobs} & \multicolumn{1}{r}{\footnotesize 31}\\
\multicolumn{1}{l}{\footnotesize 20 jobs} & \multicolumn{1}{r}{\footnotesize 21}\\
\multicolumn{1}{l}{\footnotesize 50 jobs} & \multicolumn{1}{r}{\footnotesize 31}\\
\multicolumn{1}{l}{\footnotesize 100 jobs} & \multicolumn{1}{r}{\footnotesize 38}\\
\hline
\end{tabular}
\end{center}
\end{table}

\subsection{Extension of \gate~to other domains}

Besides its application to classical emission tomography, \gate~ is potentially appropriate to simulate in-line tomography in hadrontherapy. Indeed, \gate~ possesses the ability to model the distribution of $\beta^{+}$ emitters along the beam path. In principle, this distribution can be obtained from knowledge of the nuclear cross-sections of the reactions between heavy ions and target nuclei. This distribution can then be accurately reproduced in \gate~using a voxelized source, composed of the different emitters ($^{15}$O, $^{11}$C and $^{10}$C).

\begin{figure}
\begin{center}
\includegraphics[width=.6\textwidth]{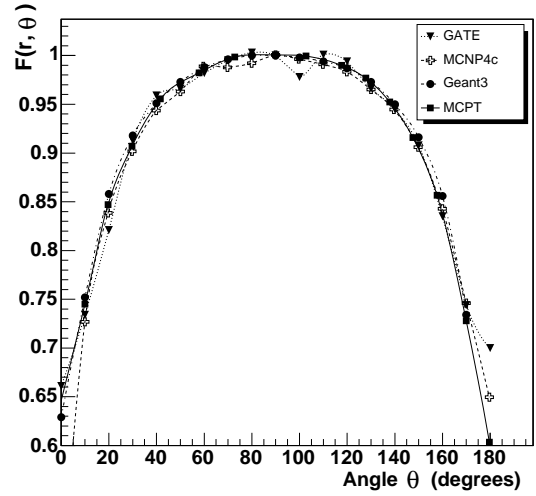}
\caption{Anisotropy functions at $r~=~2$ cm for a $^{192}$Ir brachytherapy source. The anisotropy functions $F(r,t)$ have been normalized to one at 90~$\deg$.}
\label{Anisotropy}
\end{center}
\end{figure}

Currently, Monte Carlo simulations are believed to be the most accurate method for dose calculation in radiotherapy and brachytherapy. A comparison of the computational codes in radiation dosimetry is under way (Berger \etal n.d.). In this context, the anisotropy function $F(r,\theta)$ (Nath \etal 1995) of a $^{192}$Ir brachytherapy source in water has been studied. As illustrated in Figure~\ref{Anisotropy}, anisotropy functions in water have been calculated using Geant3 (Brun \etal 1987), MCNP4C (Briesmeister 1993), MCPT (Williamson 1998), and \gate~at a radial distance of 2 cm. \gate~was in agreement with all other codes. The relative deviation concerning anisotropy functions between MCPT and \gate~simulations is less than 3\%. This indicates that \gate~has potential to perform dose calculations in brachytherapy. The challenge for \gate~is to provide detailed descriptions of local doses delivered by radiotherapy or brachytherapy treatment. It is envisioned that this can be done with voxelized phantoms in future \gate~simulations.

%% file: pmb.sec.12.tex
\section{Conclusion}
%
%

Based on \geant, \gate~is a versatile and adaptable platform for simulating PET and SPECT experiments. \gate~is appropriate for simulating conventional scanners and novel detection devices, and does not require any knowledge of C++. The \OG~collaboration, representing a large number of research groups from around the world, has publicly released this simulation toolkit after two years of software development and validation. The source code is available for download, which will enable users to make modifications to suit their particular needs. Documentation is available and user support is provided through a very active mailing list.

The future of \gate~is closely related to the future of Monte Carlo simulations in nuclear medicine. Because Monte Carlo simulations are playing an increasing role in the optimization of detector design and in the assessment of acquisition and processing protocols, we hope that \gate~will answer many of these needs. In addition, the efforts of the \OG~collaboration should ensure that \gate~will continue to evolve to become a comprehensive simulation tool at the service of the nuclear medicine community.

%% file: biblio.tex
%
%
\References 
\item[] 
Agostinelli S \etal 2003 GEANT4 - a simulation toolkit {\it Nucl. Instr. Meth.} {\bf A506} 250-303
\item[]
Assi\'e K \etal 2004 Monte Carlo simulation in PET and SPECT instrumentation using GATE {\it Nucl. Instr. Meth.} {\bf A527} 180-189
\item[]
Assi\'e K, Gardin I, Vera P, Buvat I 2004b Validation of Monte Carlo simulations of Indium 111 SPECT using GATE {\it J. Nucl. Med.} {\bf 45} 414P-415P
\item[]
Berger L \etal (nd) Intercomparison on the usage of computational codes in radiation dosimetry, ENEA/QUADOS, website: \htmladdnormallink{http:///www.enea.it:/com/ingl/default.htm}{http:///www.enea.it:/com/ingl/default.htm}
\item[]
Bielajew A F, Hirayama H, Nelson W R and Rogers D W O 1994 History, overview and recent improvements of EGS4 {\it NRCC Report} PIRS-0436,
National Research Council, Ottawa, Canada
\item[]
Briesmeister J F (Editor) 1993 MCNP - A general Monte Carlo N-particle transport code, {\it LANL Report} LA-12625-M, Los Alamos National Laboratory, Los Alamos, NM, USA
\item[]
Behrens H and J\"anecke 1969 Numerical Tables for Beta-Decay and Electron Capture, edited by Schopper H, Landolt-B\"ornstein, New Series, Group I, Vol. 4, Springer-Verlag, Berlin
\item[]
Brun R, Bruyant F, Maire M, McPherson A C, Zanarini P 1987 GEANT3 {\it Technical Report} CERN DD/EE/84-1
\item[]
Brun R, Rademakers F 1997 ROOT - An object oriented data analysis framework {\it Nucl. Instr. Meth.} {\bf A389} 81-86
\item[]
Buvat I, Castiglioni I 2002 Monte Carlo simulations in SPET and PET {\it Q. J. Nucl. Med.} {\bf 46} 48-61
\item[]
Chung Y H, Choi Y, Cho G, Choe Y S, Lee K-H, Kim B-T 2003 Optimization of dual layer phoswich detector consisting of LSO and LuYAP for small animal PET {\it in Conf. Rec. IEEE Nucl. Sci. Symp. and Med. Imag. Conf.} Portland, Oregon
\item[]
Harrison R L, Vannoy S D, Haynor D R, Gillipsie S B, Kaplan M S, Lewellen T K 1993 Preliminary experience with the photon generator module of a public-domain simulation
system for emission tomography {\it in Conf. Rec. IEEE Nucl. Sci. Symp. and Med. Imag. Conf.} San Francisco, Vol. 2, pp. 1154-1158 
\item[]
Iwata K, Greaves R G, Surko C M 1997 $\gamma$-ray spectra from positron annihilation on atoms and molecules {\it Physical Review A} {\bf 55} 3586-3604 
\item[]
James F 1990 A Review of Pseudorandom Number Generators {\it Computer Phys. Comm.} {\bf 60} 329-344
\item[]
Jan S 2002 Simulateur Monte Carlo et cam\'era \`a x\'enon liquide pour la Tomographie \`a Emission de Positons {\it PhD Thesis} Universit\'e Joseph Fourier, Grenoble, France
\item[]
Jan S \etal 2003 Monte Carlo Simulation for the ECAT EXACT HR+ system using GATE {\it submitted to IEEE Trans. Nucl. Sci.}
\item[]
Jan S, Chatziioannou A F, Comtat C, Strul D, Santin G, Trebossen R 2003b Monte Carlo simulation for the microPET P4 system using GATE {\it in Conf. Rec. HiRes 2003} Academy of Molecular Imaging, Madrid, {\it Mol. Imag. Biol.} {\bf 5} 138
\item[]
Jan S, Comtat C, Trebossen R, Syrota A 2004 Monte Carlo simulation of the MicroPET Focus for small animal {\it J. Nucl. Med.} {\bf 45} 420P
\item[]
Kawrakow I and Rogers D W O 2003 The EGSnrc code system: Monte Carlo simulation of electron and photon transport {\it NRCC Report} PIRS-701, National Research Council, Ottawa, Canada
\item[]
Kuntner C, Auffray E, Lecoq P, Pizzolotto C, Schneegans M 2002 Intrinsic energy resolution and light output of the Lu$_{0.7}$Y$_{0.3}$AP:Ce scintillator {\it Nucl. Instr. Meth.} {\bf A493} 131-136
\item[]
Lamare F, Turzo A, Bizais Y, Visvikis D 2004 Simulation of the Allegro PET system using GATE {\it to appear in Proc. of the SPIE Med. Imag. Conf.} San Diego, CA
\item[]
Lazaro D 2003  Validation de la plate-forme de simulation GATE en Tomographie d'Emission Monophotonique et application au d\'eveloppement d'un algorithme de
reconstruction 3D compl\`ete {\it PhD Thesis} Universit\'e Blaise
Pascal, Clermont Ferrand, France
\item[]
Lazaro D, Buvat I, Loudos G, Strul D, Santin G, Giokaris N, Donnarieix D, Maigne L, Spanoudaki V, Styliaris S, Staelens S, Breton V 2004 Validation of the GATE Monte Carlo simulation platform for modeling a CsI(Tl) scintillation camera dedicated to small-animal imaging {\it Phys. Med. Biol.} {\bf 49} 271-285
\item[]
{\it LGPL} (nd), GNU Lesser General Public License, Version 2.1, February 1999, Copyright (C) 1991, 1999 Free Software Foundation, Inc., 59 Temple Place, Suite 330, Boston, MA  02111-1307  USA
\item[]
Nath R, Anderson L L, Luxton G, Weaver K A, Williamson J, Meigooni A S 1995 Dosimetry of interstitial brachytherapy sources: Recommendations of the AAPM Radiation Therapy Committee Task Group No. 43, {\it Med. Phys.} {\bf 22} 209-234
\item[]
Marsaglia G, Zaman A 1987 Toward a Universal Random Number Generator, Florida State University FSU-SCRI-87-50
\item[]
Rannou F, Kohli V, Prout D, Chatziioannou A 2004 Investigation of OPET performance using GATE, a Geant4-based simulation software {\it to appear in IEEE Trans. Nucl. Sci.}
\item[]
Rey M, Simon L, Strul D, Vieira J-M, Morel C 2003 Design study of the ClearPET LSO/LuYAP phoswich detector head using GATE {\it in Conf. Rec. HiRes 2003} Academy of Molecular Imaging, Madrid, {\it Mol. Imag. Biol.} {\bf 5} 119
\item[]
Rogers D W O, Faddegon B A, Ding G X, Ma C-M, Wei J, and Mackie T R 1995 BEAM: A Monte Carlo code to simulate radiotherapy treatment units,
{\it Med. Phys.} {\bf 22} 503 - 524
\item[]
Santin G, Strul D, Lazaro D, Simon L, Krieguer M, Vieira Martins M, Breton V, Morel C 2003 GATE: A GEANT4-based simulation platform for PET and SPECT integrating movement and time management {\it IEEE Trans. Nucl. Sci.} {\bf 50} 1516-1521
\item[]
Schmidtlein C R, Nehmeh S A, Bidaut L M, Erdi Y E, Humm J L, Amols H I, Kirov A S 2004 Validation of GATE Monte Carlo simulations for the GE Advance PET scanner {\it J. Nucl. Med.} {\bf 45} 409P-410P
\item[]
Simon L, Strul D, Santin G, Krieguer M, Morel C 2004 Simulation of time curves in small animal PET using GATE {\it Nucl. Instr. Meth.} {\bf A527} 190-194
\item[]
Staelens S, Strul D, Santin G, Koole M, Vandenberghe S, D'Asseler Y, Lemahieu I, Van de Walle R 2003 Monte Carlo simulations of a scintillation camera using GATE: validation and application modeling {\it Phys. Med. Biol.} {\bf 48} 3021-3042
\item[]
Staelens S, D'Asseler Y, Vandenberghe S, Koole M, Lemahieu I, Van de Walle R, 2004 A three-dimensional theoretical model incorporating spatial detection uncertainty in continuous detector PET {\it to appear in Phys. Med. Biol.}
\item[]
Staelens S, Santin G, Vandenberghe S, Strul D, Koole M, D'Asseler Y, Lemahieu I, Van de Walle R 2004b Transmission imaging with a moving point source: influence of crystal thickness and collimator type {\it to appear in IEEE Trans. Nucl. Sci.}
\item[] 
Staelens S, Koole M, Vandenberghe S, D'Asseler Y, Lemahieu I, Van de Walle R 2004c The geometric transfer function for a slat collimator mounted on a strip detector {\it to appear in IEEE Trans. Nucl. Sci.}
\item[]
Staelens S, Vandenberghe S, De Beenhouwer J, De Clercq S, D'Asseler Y, Lemahieu I, Van de Walle R 2004d A simulation study comparing the imaging performance of a solid state detector with a rotating slat collimator versus a traditional scintillation camera {\it to appear in Proc. of the SPIE Med. Imag. Conf.} San Diego, CA
\item[]
Strul D 2001 Preliminary specifications of a Geant4-based framework for nuclear medicine simulations {\it ClearPET Technical Report} University of Lausanne, Switzerland
\item[]
Strul D 2001b Specification of a Geant4-based nuclear medicine simulation framework {\it ClearPET Technical Report} University of Lausanne, Switzerland
\item[]
Strul D, Santin G, Lazaro D, Breton V, Morel C 2003 GATE (GEANT4 Application for Tomographic Emission): a PET/SPECT general-purpose simulation platform {\it Nucl. Phys. B (Proc. Suppl.)} {\bf 125C} 75-79
\item[]
Traore M, Hill D 2001 The use of random number generation for stochastic distributed simulation: application to ecological modeling {\it in Conf. Rec. 13$^{th}$ European Simulation Symposium} Marseille, pp. 555-559
\item[]
von Neumann J 1951 Various techniques in connection with random digits-Monte Carlo methods {\it Nat. Bureau Standards}, AMS 12, pp. 36-38 
\item[]
Williamson J F 1998 Monte Carlo simulation of photon transport phenomena, in Monte Carlo Simulation in the Radiological Sciences, edited by R. L. Morin, CRC Press, Boca Raton, pp. 53-102
\item[] 
Ziemons K \etal 2003 The ClearPET LSO/LuYAP phoswich scanner: a high performance small animal PET system {\it in Conf. Rec. IEEE Nucl. Sci. Symp. and Med. Imag. Conf.} Portland, Oregon
\endrefs